\documentclass[jcp,aip]{revtex4-1}
\usepackage{graphicx, hyperref, multirow, fontenc}

\begin{document}

\title{First-principles predicted low-energy structures of NaSc(BH$_4$)$_4$}

\author{Huan Doan Tran}
\affiliation{Department of Physics, Universit\"{a}t Basel, Klingelbergstrasse 82, 4056 Basel, Switzerland}
\altaffiliation{Present address of HDT: Institute of Materials Science, University of Connecticut, 97 North Eagleville Rd., Unit 3136, Storrs, CT 06269-3136, USA; email: huan.tran@uconn.edu}
\author{Maximilian Amsler}
\affiliation{Department of Physics, Universit\"{a}t Basel, Klingelbergstrasse 82, 4056 Basel, Switzerland}
\author{Silvana Botti}
\affiliation{Universit\'e de Lyon, F-69000 Lyon, France and LPMCN, CNRS, UMR 5586, Universit\'e Lyon 1, F-69622 Villeurbanne, France}
\author{Miguel A. L. Marques}
\affiliation{Universit\'e de Lyon, F-69000 Lyon, France and LPMCN, CNRS, UMR 5586, Universit\'e Lyon 1, F-69622 Villeurbanne, France}
\author{Stefan Goedecker}
\email{stefan.goedecker@unibas.ch}
\affiliation{Department of Physics, Universit\"{a}t Basel, Klingelbergstrasse 82, 4056 Basel, Switzerland}

\date{\today}

\begin{abstract}
According to previous interpretations of experimental data,  
sodium-scandium double-cation borohydride NaSc(BH$_4$)$_4$ crystallizes in the crystallographic space group $Cmcm$ where each sodium (scandium) atom is surrounded by six scandium (sodium) atoms. A careful investigation of this phase based on \textit{ab initio} calculations indicates that the structure is dynamically unstable and gives rise to an energetically and dynamically more favorable phase with $C222_1$ symmetry and nearly identical x-ray diffraction pattern. By additionally performing extensive structural searches with the minima-hopping method we discover a class of new low-energy structures exhibiting a novel structural motif in which each sodium (scandium) atom is surrounded by four scandium (sodium) atoms arranged at the corners of either a rectangle with nearly equal sides or a tetrahedron. These new phases are all predicted to be insulators with band gaps of $7.9-8.2$ eV. Finally, we estimate the influence of these structures on the hydrogen-storage performance of NaSc(BH$_4$)$_4$.
\end{abstract}


\maketitle

\section{Introduction} \label{sec:intro}
The imminent shortage of fossil resources has sparked an intense search for alternative energy sources in the last decades, and hydrogen has been considered as a promising candidate due to its clean reaction with oxygen. However, engineering suitable solid hydrogen storage media with high energy density and appropriate hydrogenation/dehydrogenation properties has proven to be a challenging task. Aside from simple metal hydrides like MgH$_2$, LiH$_2$ or LiAlH$_4$, other materials have been recently proposed with increasing complexity. The successful synthesis of double-cation borohydrides such as Li/K, \cite{Nickels:LiK} Li/Ca, \cite{Fang:LiCa} Li/Sc, \cite{Hagemann:LiSc} K/Sc, \cite{Cerny:KSc} Na/Al, \cite{Lindemann:NaAl} K/Mn and K/Mg, \cite{Schouwink_KMnKMg} K/Y, \cite{Jaron:KY} Li/Zn, Na/Zn, and K/Zn \cite{RavnsbaekLiNa_Zn, Cerny10:LiNa, Ravnsbaek:LiZn}  borohydrides has drawn much attention as potential candidates for hydrogen-storage.~\cite{schlapbach}

NaSc(BH$_4$)$_4$, a double-cation borohydride of sodium and scandium which contains $12.67$ wt.\% hydrogen, was recently synthesized \cite{Cerny:NaSc} by ball milling of NaBH$_4$ and ScCl$_3$
\begin{equation}
\label{Eq:synth}
\rm
4NaBH_4+2ScCl_3 \to NaSc(BH_4)_4+Na_3ScCl_6.
\end{equation}
Below the melting temperature ($\simeq$ 410 K), NaSc(BH$_4$)$_4$ was experimentally determined to belong to the $Cmcm$ crystallographic space group (no. 63). \cite{Cerny:NaSc} In this phase,  NaSc(BH$_4$)$_4$ is an ionic crystal of which each Na$^+$ cation is octahedrally coordinated to six [Sc(BH$_4$)$_4$]$^-$ complex anions. The scandium atom of each [Sc(BH$_4$)$_4$]$^-$ complex is surrounded by four [BH$_4$] tetrahedra, forming a nearly ideal tetrahedron with a Sc-B distance between 2.27\AA~ and 2.50\AA. Above 410 K, NaSc(BH$_4$)$_4$ decomposes while releasing hydrogen in two rapid steps between 440~K and 490~K, and between 495~K and 540~K, respectively. \cite{Cerny:NaSc} Details on these decomposition steps have not been experimentally determined yet. \cite{Cerny:NaSc}

Soon after the experimental synthesis of NaSc(BH$_4$)$_4$, several theoretical studies were carried out. Huang {\it et al.} \cite{Huang:NaSc} reported that the $Cmcm$ phase of NaSc(BH$_4$)$_4$ is an insulator with a band gap of $E_{\rm g}=5.055$ eV. The following single-step decomposition reaction with a release of 9.3 wt.\% hydrogen gas was theoretically predicted by Kim~\cite{Kim12:NaSc, Kim13:NaSc} to maximize the gain of Landau free energy
\begin{equation}\label{Eq:decomp}
\rm NaSc(BH_4)_4 \to ScB_2+\frac{3}{4}NaBH_4+\frac{1}{8}Na_2B_{10}H_{10}+\frac{47}{8}H_2.
\end{equation}
The pressure $P_{\rm H_2}$ at which this hypothetical reaction occurs is related to the temperature $T$  through
 \cite{Alapati07a, Alapati07b}
\begin{equation}\label{Eq:vHoff}
\frac{P_{\rm H_2}}{P_0}=\exp\left[-\frac{\Delta G(T)}{RT}\right],
\end{equation}
where $\Delta G(T)$ is the Gibbs free energy change of the reaction (\ref{Eq:decomp}) at $T$, $P_0={\rm 1atm}$, and $R$ is the gas constant. The entropic contribution from the hydrogen (gas) product to $\Delta G(T)$ is determined by the Shomate equation for which the coefficients can be taken from the database of the National Institute of Standards and Technology (NIST). \cite{NIST} Other contributions to $\Delta G(T)$ can be calculated at the density functional theory (DFT) \cite{dft1, dft2} level. Kim predicted \cite{Kim12:NaSc, Kim13:NaSc} that, at $P_{\rm H_2}=100$ bar, the proposed reaction (\ref{Eq:decomp}) occurs at temperatures between 14 K and 223 K. The wide range of the predicted temperature $T$ is due to the DFT uncertainty in calculating $\Delta G(T)$ and comparing with the measured data. Although the temperature calculated for the hypothetical reaction (\ref{Eq:decomp}) strongly differs from the experimental decomposition temperature ($440-540$ K),~\cite{Cerny:NaSc} relation (\ref{Eq:vHoff}) indicates that at a given temperature $T$, the required H$_2$ pressure $P_{\rm H_2}$ is sensitive to the free energy change $\Delta G(T)$ since $RT$ is small (at the room temperature $T=298~\rm K$, $RT \simeq 2.5 \rm kJ/mol~{\rm H}_2$). Therefore, a change in $P_{\rm H_2}$ may be expected for a new low-energy phase of NaSc(BH$_4$)$_4$ even with a small difference in energy with respect to the $Cmcm$ phase.

In this work we carefully study the $Cmcm$ structure of NaSc(BH$_4$)$_4$ with \textit{ab initio} calculations, showing that it is dynamically unstable. By following the imaginary frequency phonon modes, a dynamically stable structure with $C222_1$ symmetry is predicted. Furthermore, we report a class of new low-energy structures predicted by the minima-hopping method \cite{Goedecker:MHM, Amsler:MHM} (MHM) for this material. Finally we discuss the influence of small changes in the free energy $\Delta G(T)$ of these new structures on the hydrogen pressure $P_{\rm H_2}$ in reaction (\ref{Eq:decomp}).

\section{Computational methods}
We used the projector augmented wave formalism \cite{PAW} as implemented in the {\it Vienna Ab Initio Simulation Package} ({\sc vasp}) \cite{vasp1,vasp2,vasp3,vasp4} to perform all the DFT-based first-principles calculations. The valence electron configurations of sodium, scandium, boron, and hydrogen were $2p^6 3s^1$, $3s^2 3p^6 4s^2 3d^1$, $2s^2 2p^1$ and $1s^1$, respectively. The DFT total energy $E_{\rm DFT}$ was calculated with Monkhorst-Pack $\bf k$-point meshes \cite{monkhorst} with sizes of either $5 \times 5 \times 5$ or $7 \times 7 \times 7$, depending on the volume of the simulation cell, and a plane wave kinetic energy of 450 eV. The convergence of electronic self-consistent (SC) calculations is assumed when the total energy change between two consecutive steps is smaller then $10^{-6}$ eV. Atomic and cell variables were simultaneously relaxed until the residual forces were smaller than $10^{-2}$~eV/\AA.

For systematically exploring the low-energy landscape corresponding to NaSc(BH$_4$)$_4$, we employed the minima-hopping (MH) method.~\cite{Goedecker:MHM,Amsler:MHM} Being needed by the MH algorithm, the total energies $E_{\rm DFT}$, the forces and the stresses of the examined structures are calculated at the DFT level by {\sc vasp} in this work. They are then used to perform a number of consecutive short molecular dynamics steps, driving the system out of the current minimum at which the system is trapped. At the end of the molecular dynamics stage, the obtained structure is optimized by ordinary (local) geometry relaxations until the convergence criteria described above are met. By choosing the initial velocities of the molecular dynamics trajectories approximately along soft mode directions, the likelihood of escaping the current minimum and landing at a lower-energy minimum can be significantly improved.\cite{Goedecker:MHM} In addition, several build-in feedback mechanisms are implemented, minimizing the possibility of revisiting the already explored minima and  enhancing the efficiency of the method.\cite{Goedecker:MHM,Amsler:MHM} Because the MH method requires no constrain, unknown structural motifs can be explored. The reliability of this method was shown in many applications that were recently reported. \cite{hellmann_2007, roy_2009, bao_2009, willand_2010, de11, amsler_crystal_2011, Livas, amsler12, HuanZn, HuanAlanates}

The super-cell approach implemented in {\sc Phonopy} \cite{phonopy, phonopy_sc} was used to analyze the phonon frequency spectrum and investigate the dynamical stability of the structures examined. In this approach, the phonon frequencies were calculated from the dynamical matrix of which the force constants were evaluated in {\sc vasp}. Sufficiently high convergent criteria are required for these force calculations. In particular, the electronic SC loops are terminated when two consecutive steps differ by less than $10^{-8}$ eV in energy while the residual forces on the atoms have to be smaller than $10^{-4}$~eV/\AA. The longitudinal optical/transverse optical (LO/TO) splitting was not taken into account because its effects on dynamical properties were reported to be negligible for a wide variety of hydrides. \cite{hector07,herbst10} 

\section{Effects of exchange-correlation functionals and vdW interactions} \label{sec:validate}

It is well established that the lack of long-range van der Waals (vdW) interactions in DFT, in many cases, may significantly affect its accuracy  when investigating soft matter and molecular crystals. Studying magnesium borohydride Mg(BH$_4$)$_2$, Bil {\it et al.} \cite{Bil:vdW} pointed out that DFT calculations with the Perdew-Burke-Ernzerhof (PBE) exchange-correlation (XC) functional artificially favors structures with unusually low densities. \cite{OzolinsMgBH4, VossMgBH4} The reason for this observation can probably be traced back to the exceptionally complicated experimental structure of Mg(BH$_4$)$_2$ which has 330 atoms per cell and is composed of several {\it neutral} sub-structures (frameworks) held together by dispersion interactions. \cite{DaiMgBH4,FilinchukMgBH4} Because conventional DFT calculations do not well capture these interactions, a non-local density functional such as vdW-DF \cite{Dion-DF, Thonhauser-DF, Roman-DF} was suggested \cite{Bil:vdW} to be more suitable. Similarly, the experimental structures of some double-cation borohydrides studied in Ref. \onlinecite{Huan:Mixed}, e.g., LiZn$_2$(BH$_4$)$_5$, NaZn(BH$_4$)$_3$, and NaZn$_2$(BH$_4$)$_5$, are composed of two neutral, inter-penetrated frameworks.\cite{RavnsbaekLiNa_Zn} For these materials, PBE calculations predict very low-density structures to be thermodynamically stable while, at the same time, produce large deviations when optimizing their complicated experimental structures. \cite{Huan:Mixed} By taking into account the vdW interactions via the non-local density functional vdW-DF2, \cite{vdW-DF2} the optimized geometries agree better with the experimental data. It is worth noting that for the borohydrides with simple ionic crystalline structures, e.g., LiBH$_4$, NaBH$_4$, and KBH$_4$, the calculated results with vdW-DF2 are not always better than those with PBE, \cite{Huan:Mixed} indicating that, presumably, the dispersion interactions play a minor role in the crystalline materials dominated by ionic and covalent bonds. Clearly, the performance of a vdW treatment, e.g., vdW-DF2, is material-dependent, i.e., it depends on the nature of the interactions in the materials examined.

\begin{table*}[t]
\begin{center}
\caption{Structural parameters from the DFT optimization of the $Cmcm$ structure (left) and $C222_1$ structure (right) of NaSc(BH$_4$)$_4$, calculated with and without vdW interactions and given in \AA. The differences $\Delta$ between the calculated results and the experimental data are given in \%. The space group of the structure was not changed by the optimization. Experimental data are taken from Ref. \onlinecite{Cerny:NaSc}.} \label{table:test}
\begin{tabular}{l rrrrrrr r r | r r r r r r r r r r r r r r}
\hline
\hline
& \multicolumn{2}{c}{PW91} && \multicolumn{2}{c}{PBE} && \multicolumn{2}{c}{vdW-DF2} &&& \multicolumn{2}{c}{PW91} && \multicolumn{2}{c}{PBE} && \multicolumn{2}{c}{vdW-DF2} && \multirow{2}{*}{Exp.}\\
\cline{2-3} \cline{5-6} \cline{8-9} \cline {12-13} \cline{15-16} \cline{18-19}
& DFT & $\Delta$(\%) & & DFT & $\Delta$(\%) & & DFT & $\Delta$(\%) & & & DFT & $\Delta$(\%) & &DFT & $\Delta$(\%)&& DFT & $\Delta$(\%) &\\
\hline
$a$(\AA)    &8.109  &$-0.7$ && 8.122  & $-0.6$ &&   8.088  &$-1.0$ &&&8.312&$1.7$&&8.318& $1.8$ &&8.131&$-0.4$&& 8.170 \\
$b$(\AA)    &11.900 &0.2    && 11.904 & $ 0.2$ &&   11.765 &$-0.9$ &&&11.821&$-0.5$&&11.827&$-0.4$ &&11.720&$-1.3$&&11.875 \\
$c$(\AA)    &8.967  &$-0.6$ && 8.973  & $-0.5$ & &   8.640  &$-4.2$ &&&9.111&$1.0$&& 9.117&$1.1$ &&8.628&$-4.3$&& 9.018   \\
$V$(\AA$^3$)&865.4  &$-1.1$ && 867.6  & $-0.8$ & & 822.2  &$-6.0$ &&&895.2&$2.3$&&896.9&$2.5$&&822.2&$-6.0$&& 874.9 \\
\hline
\hline
\end{tabular}
\end{center}
\end{table*}

For NaSc(BH$_4$)$_4$, we have tested our calculations with the available vdW methods and the commonly used XC functionals implemented in {\sc vasp} by fully optimizing the $Cmcm$ structure and studying $\Delta V$, the differences of the optimized cell volume $V$, with respect to the experimental value. The best results are summarized in Table \ref{table:test} while the full data set is given in the Supplemental Material.\cite{supplement} For vdW-DF2, $\Delta V = -6.0\%$, while for PBE and Perdew-Wang (PW91) XC functionals, $\Delta V$ are much smaller and almost equal ($-0.8 \%$ with PBE and $-1.1 \%$ with PW91). Our calculations also indicate that the unit cell of the $Cmcm$ structure was strongly distorted with $\Delta V = -9.0\%$ when employing the PBEsol functional~\cite{PBEsol} and with $\Delta V = -17.3\%$ when the local density approximation (LDA) is used. Two other implementations of the vdW, i.e., DFT-D2 \cite{DFT-D2} and DFT-TS \cite{DFT-TS, vdW-TS-Bucko} are known \cite{vdW-TS-Bucko, Bucko_commun} to significantly underestimate the lattice parameters of ionic crystals. We actually obtained very large reductions of the unit cell volume with $\Delta V = -17.5\%$ for DFT-D2 and $\Delta V = -16.5\%$ for DFT-TS. 

The relatively large volume change $\Delta V=-6.0 \%$ obtained with vdW-DF2, five times larger than the very good (small) values of $\Delta V$ obtained by PBE and PW91, may indicate the minor role of the dispersion interactions in NaSc(BH$_4$)$_4$. To further examine this marginal case, various characteristic distances and angles of the $Cmcm$ structure optimized with PW91, PBE, and vdW-DF2  were compared with the experimental data. \cite{supplement} The optimized geometries of the complex [Sc(BH$_4$)$_4]^{-}$ anions are almost the same and agree very well with the experimental data, suggesting that the short-range interactions are well captured by our calculations. The long distances between any Na$^+$ cation and the six surrounding [Sc(BH$_4$)$_4]^{-}$ anions (the distance between Na and Sc were actually measured) are however {\it isotropically} reduced by $\simeq 2\%$ in the structure optimized with vdW-DF2, indicating that in this ionic crystalline structure, the long-range interactions are systematically overestimated. \cite{supplement} As the calculations with PBE and PW91 reduce the Na-Sc lengths by no more than $0.7 \%$ without changing the relevant angles, these functionals are clearly more favorable than vdW-DF2. Although PBE was used throughout this work, results with vdW-DF2 are also carefully referred when necessary.

\section{Low-energy structures of sodium/scandium borohydride}

\begin{figure}[t]
  \begin{center}
    \includegraphics[width=8.25 cm]{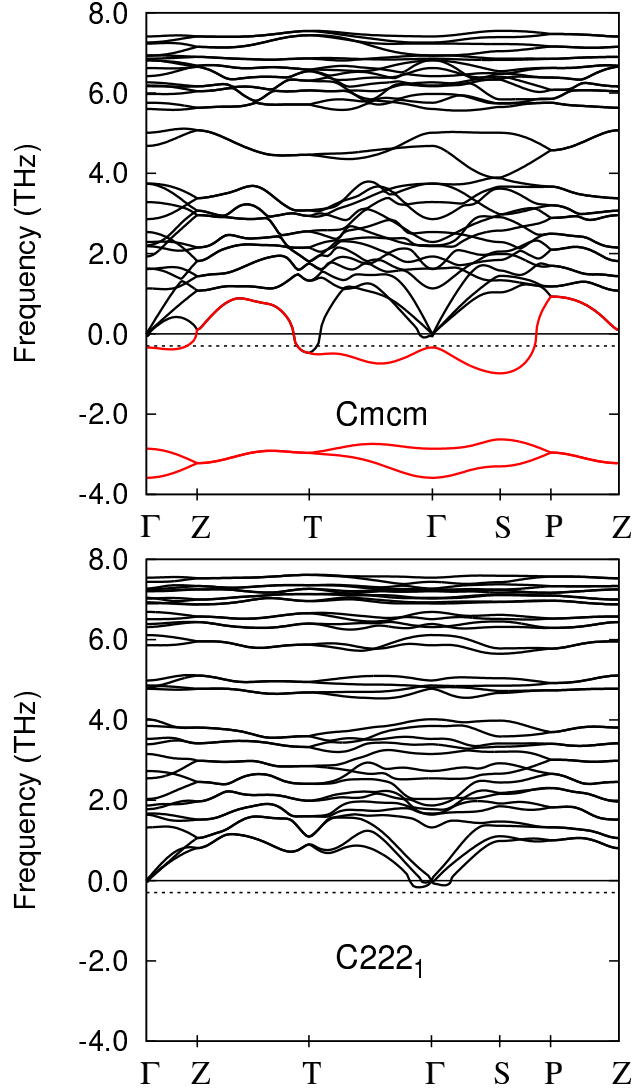}
  \caption{(Color online) Phonon band structures of the $Cmcm$ (top panel) and $C222_1$ (bottom panel) structures of NaSc(BH$_4$)$_4$. Three imaginary phonon modes of the $Cmcm$ structure are shown in red. For convenience, bands with imaginary frequencies are shown in this figure as those with {\it negative} frequencies of the same module. Dotted lines indicate the lower bound corresponding to the errorbar of $\sim 0.3$ THz due to the numerically unresolved translational invariance in calculations of the XC energies. \cite{phonon_error,VossMgBH4}} \label{fig:band}
  \end{center}
\end{figure}

\begin{figure}[t]
  \begin{center}
    \includegraphics[width=6 cm]{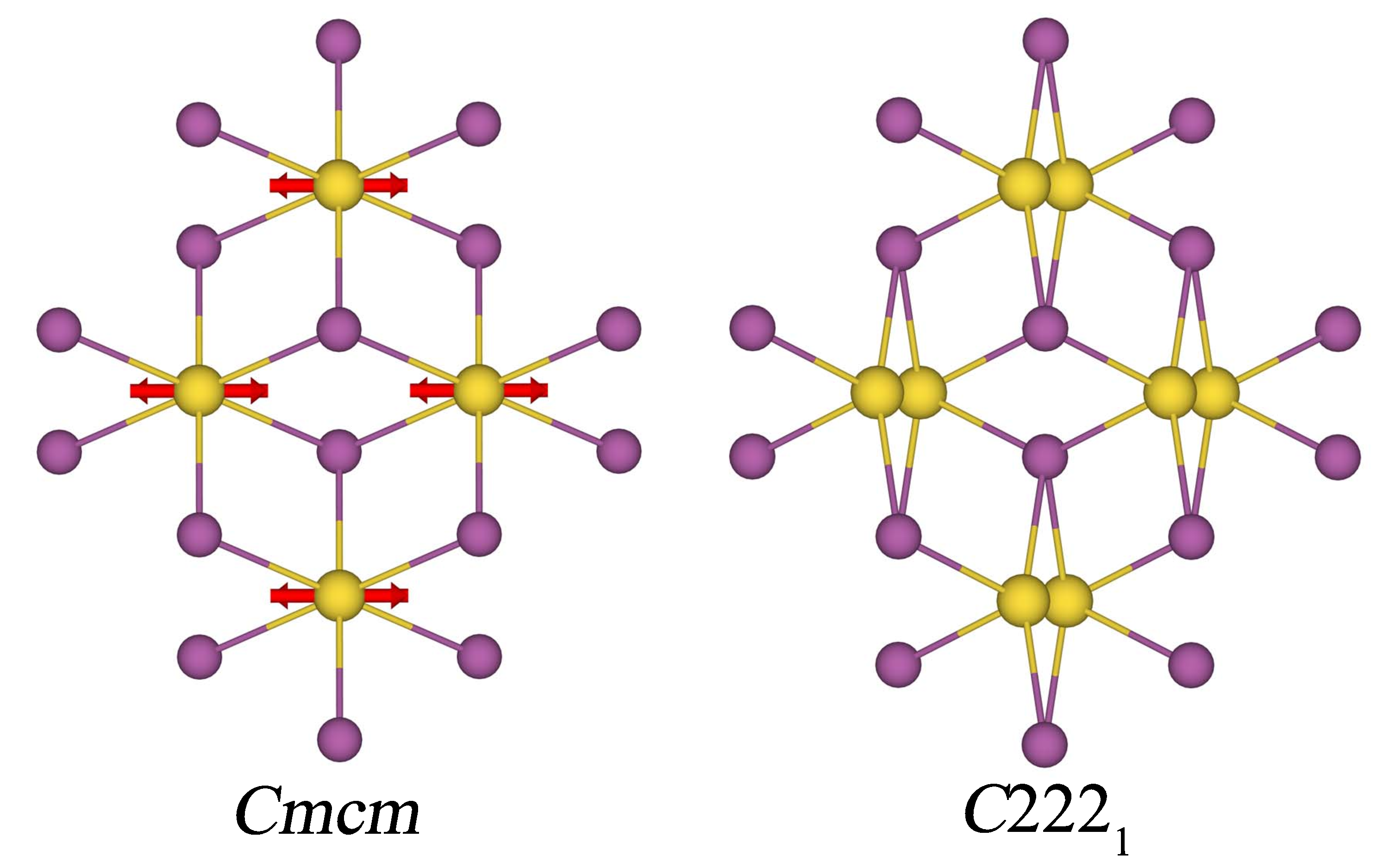}
  \caption{(Color online) Top view (along the $z$ axis) of the $Cmcm$ and $C222_1$ structures of NaSc(BH$_4$)$_2$. Yellow and purple spheres represent sodium and scandium atoms. Red arrows indicate schematically the eigenvector of the unstable mode following which the centrosymmetry of the $Cmcm$ structure is broken, and the $C222_1$ structure is formed.} \label{fig:deform}
  \end{center}
\end{figure} 

\subsection{The $Cmcm$ and $C222_1$ structures}\label{sec:Cmcm}

We performed phonon calculations for the $Cmcm$ structure and show the result in the top panel of Fig. \ref{fig:band}. We found that, at the DFT level, the $Cmcm$ phase is dynamically unstable with three imaginary phonon modes emerging at the $\Gamma$ point, two of them have large imaginary frequencies, roughly $3i$THz, throughout the whole Brillouin zone. One of these two modes describes the instability of the Na atomic positions, which are the inversion centers, and the other indicates that the BH$_4$ tetrahedra must also be stabilized. By following these modes, the BH$_4$ tetrahedra are slightly rearranged while the centrosymmetry is broken by driving the Na atoms out of the inversion centers. The lowest possible structure obtained by this process belongs to the $C222_1$ crystallographic space group (no. 20) and is lower in energy than the $Cmcm$ structure by 3.8 kJ mol$^{-1}$f.u.$^{-1}$. We note that although comparing with experimental data, DFT calculations for the free energy change of a reaction may suffer an uncertainty of $\sim 10$kJ/mol, \cite{Kim12:NaSc, Kim13:NaSc,Alapati07a,Alapati07b} such an certainty is expected to be canceled out when $E_{\rm DFT}$ calculated at the same computational conditions for similar structures are compared. Therefore, the energetic ordering of the structures examined throughout this work, specifically $Cmcm$ and $C222_1$, can be determined with essentially no uncertainty. We then calculated the phonon band structure of the $C222_1$ structure and show it in the bottom panel of Fig. \ref{fig:band}. Within an errorbar of $\sim 0.3$ THz due to the translational invariance breaking in XC energies calculations,\cite{phonon_error,VossMgBH4} the $C222_1$ structure is found to be dynamically stable. We illustrate the transformation from the $Cmcm$ phase to the $C222_1$ phase in Fig. \ref{fig:deform} while detailed information of the $C222_1$ structure is given in the Supplemental Material.\cite{supplement}

To further examine this conclusion, we performed additional phonon calculations for the $Cmcm$ and $C222_1$ structures with vdW-DF2 and LDA, giving the phonon band structures shown in the Supplemental Material. \cite{supplement} We found that the vdW-DF2 and LDA results agree well with the PBE results. In particular, the $Cmcm$ structure is dynamically unstable, the $C222_1$ structure is dynamically stable, and the obtained phonon band structures are very similar to those obtained with PBE (see Fig. \ref{fig:band} and the Supplemental Material \cite{supplement}). In addition, the $C222_1$ structure is energetically more stable than the $Cmcm$ structure by 1.3 kJ mol$^{-1}$f.u.$^{-1}$ with vdW-DF2 and by 0.6 kJ mol$^{-1}$f.u.$^{-1}$ with LDA. It is worth to emphasize that in the calculations of phonon band structures at the level of density functional theory, specifically by the direct method, \cite{phonopy, phonopy_sc} energy and forces are evaluated for structures with very small perturbations (at order of 0.01\AA), implying a high degree of error cancellation. Very often, the dynamical stabilities determined by first-principle calculations for various crystalline materials were found to be in excellent agreement with experiments. \cite{Schmidt:Phonon,Yildirim:MgB2Phonon} In this work, that the results obtained with PBE, vdW-DF2, and LDA are highly consistent indicates that the stabilities of the $Cmcm$ and $C222_1$ structures were correctly determined.

\begin{figure}[t]
  \begin{center}
    \includegraphics[width=8.25 cm]{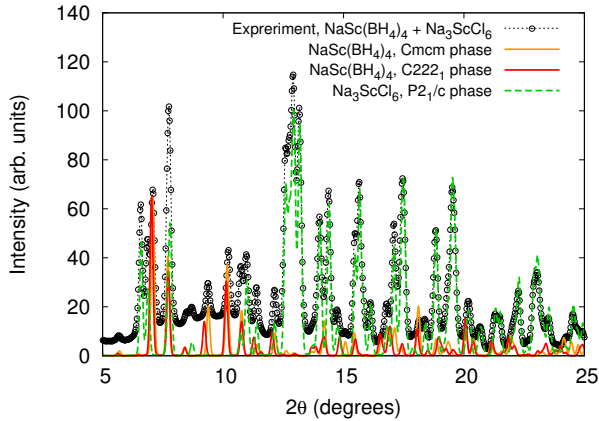}
  \caption{(Color online) Powder XRD patterns of the $P2_1/c$ structure of Na$_3$ScCl$_6$ and the $Cmcm$ and $C222_1$ structures of NaSc(BH$_4$)$_4$ which were simulated at the wavelength 0.66863 \AA~. The experimental XRD pattern of the mixture of NaSc(BH$_4$)$_4$ and Na$_3$ScCl$_6$ which was measured at the same wavelength by Ref. \onlinecite{Cerny:NaSc}, is also shown.} \label{fig:xrd}
  \end{center}
\end{figure}

In Ref. \onlinecite {Cerny:NaSc}, a powder x-ray diffraction (XRD) pattern was measured for a mixture of NaSc(BH$_4$)$_4$ and Na$_3$ScCl$_6$, according to the synthesis reaction (\ref{Eq:synth}). To further analyze the $C222_1$ structure, we used the {\sc fullprof} package \cite{fullprof} to calculate the XRD patterns of the $Cmcm$ and $C222_1$ structures of NaSc(BH$_4$)$_4$ and the $P2_1/c$ structure of Na$_3$ScCl$_6$. The simulated XRD patterns are shown in Fig. \ref{fig:xrd}, demonstrating that the $C222_1$ and $Cmcm$ structures equally explain the available experimental XRD data. In particular, although the experimental XRD pattern is dominated by peaks of Na$_3$ScCl$_6$ as mentioned by Ref. \onlinecite{Cerny:NaSc}, the simulated XRD patterns of the $Cmcm$ and $C222_1$ structures are nearly identical and resolve all the major peaks identified to belong to NaSc(BH$_4$)$_4$.

Surprisingly, the $C222_1$ structure was already considered in Ref. \onlinecite{Cerny:NaSc} as a candidate for the low-energy structure of NaSc(BH$_4$)$_4$ before being superseded by the $Cmcm$ structure. The first order phase transition from $Cmcm$ to $C222_1$ by following the soft modes breaks the centrosymmetry, leaving the overall structural motif and atomic coordination intact. The unit cell volume of the $C222_1$ structure is $897$ \AA$^3$, which is 2.5\% larger than that of the $Cmcm$ structure ($875$ \AA$^3$). We show in Table \ref{table:test} the optimized cell parameters of the $C222_1$ structure with PBE, PW91, and vdW-DF2, revealing a good agreement between the PBE and PW91 results with the experimental data. In summary, we can confidently conclude that the $C222_1$ is in fact the experimentally observed structure of NaSc(BH$_4$)$_4$.

\subsection{Minima-hopping predicted low-energy structures}
Aside from the phonon-mode-following approach, we conducted unconstrained, systematic structural searches with several MHM simulations for one and two formula units (22 or 44 atoms) per simulation cell to explore the low energy configurations of NaSc(BH$_4$)$_4$. Thereby we discovered a new class of structures with a different structural motif compared to the $C222_1$ phase. Four new structures of this class with  $C2$ (no. 5), $Cc$ (no. 9), $P1$ (no. 1), and $I222$ (no. 23) symmetry were found to have lower total energy $E_{\rm DFT}^{\rm PBE}$ than the $Cmcm$  phase (see Table \ref{table_energy}), while only three of them ($C2$, $Cc$, and $P1$) are thermodynamically more stable than the $C222_1$ structure by up to 5.07 kJ mol$^{-1}$f.u.$^{-1}$. Calculations for the total energy $E_{\rm DFT}^{\rm vdW}$ with vdW-DF2, however, indicate that these new structures are less stable than the $C222_1$ structure by up to 5.27 kJ mol$^{-1}$f.u.$^{-1}$. As discussed in Sec. \ref{sec:validate}, we believe that calculations with vdW-DF2 overestimate the long-range interactions in NaSc(BH$_4$)$_4$, and this could be the reason for these small energy differences. Treating this result with caution, we are able to conclude that the predicted structures are energetically competing with the experimentally-synthesized $C222_1$ structure of NaSc(BH$_4$)$_4$.

\begin{figure}[b]
  \begin{center}
    \includegraphics[width=8.25 cm]{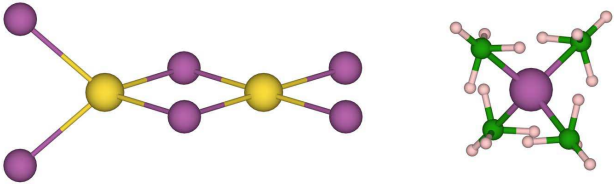}
  \caption{(Color online) Coordination of sodium atoms (left) and the geometry of [Sc(BH$_4$)$_4$]$^-$ complex anion (right) in the novel structural motifs of the low-energy NaSc(BH$_4$)$_4$ phases. Yellow, purple, green, and pink spheres represent sodium, scandium, boron, and hydrogen atoms.} \label{fig:coordination}
  \end{center}
\end{figure}

\begin{table}[b]
\begin{center}
\caption{Summary of the low-energy structures of NaSc(BH$_4$)$_4$. DFT energy $E^{\rm PBE}_{\rm DFT}$, $E^{\rm vdW}_{\rm DFT}$ and free energy $F_T$ at temperature $T$  are given in kJ mol$^{-1}$f.u.$^{-1}$ with respect to the $C222_1$  structure. Energy band gap $E_{\rm g}^{\rm GW}$ is given in units of eV while the minimum H$_2$ pressure $P_{\rm H_2}^{\rm min}(298{\rm K})$ is given in bar.} \label{table_energy}
\begin{tabular}{lrrrrrr}
\hline
\hline
Structure & $E_{\rm DFT}^{\rm PBE}$ & $F_{\rm 0K}$ & $F_{\rm 298K}$ & $E_{\rm DFT}^{\rm vdW}$ & $E_{\rm g}^{\rm GW}$&$P_{\rm H_2}^{\rm min}(298{\rm K})$\\
\hline
$C2$ (5)   & $-5.07$ & $-4.06$ & $-4.99$  &$4.75$& $8.17$ & $546$ \\
$Cc$ (9)   & $-4.84$ & $-3.73$ & $-4.67$  &$5.27$& $8.20$ & $558$\\
$P1$ (1)   & $-1.14$ & $-0.54$ & $-0.97$  &$4.05$& $7.94$ & $720$\\
$C222_1$ (20) & $0.00$  & $0.00$ & $0.00$ &$0.00$& $7.85$ &$770$ \\
$I222$ (23) & $1.44$ & $1.56$ & $0.83$    &$4.08$ & $7.91$& $815$\\
$Cmcm$ (63) & $3.80$  & $-$ & $-$         &$1.35$ & $8.10$ & $1000$\\
\hline
\hline
\end{tabular}
\end{center}
\end{table}

\begin{figure*}[t]
  \begin{center}
    \includegraphics[width=15 cm]{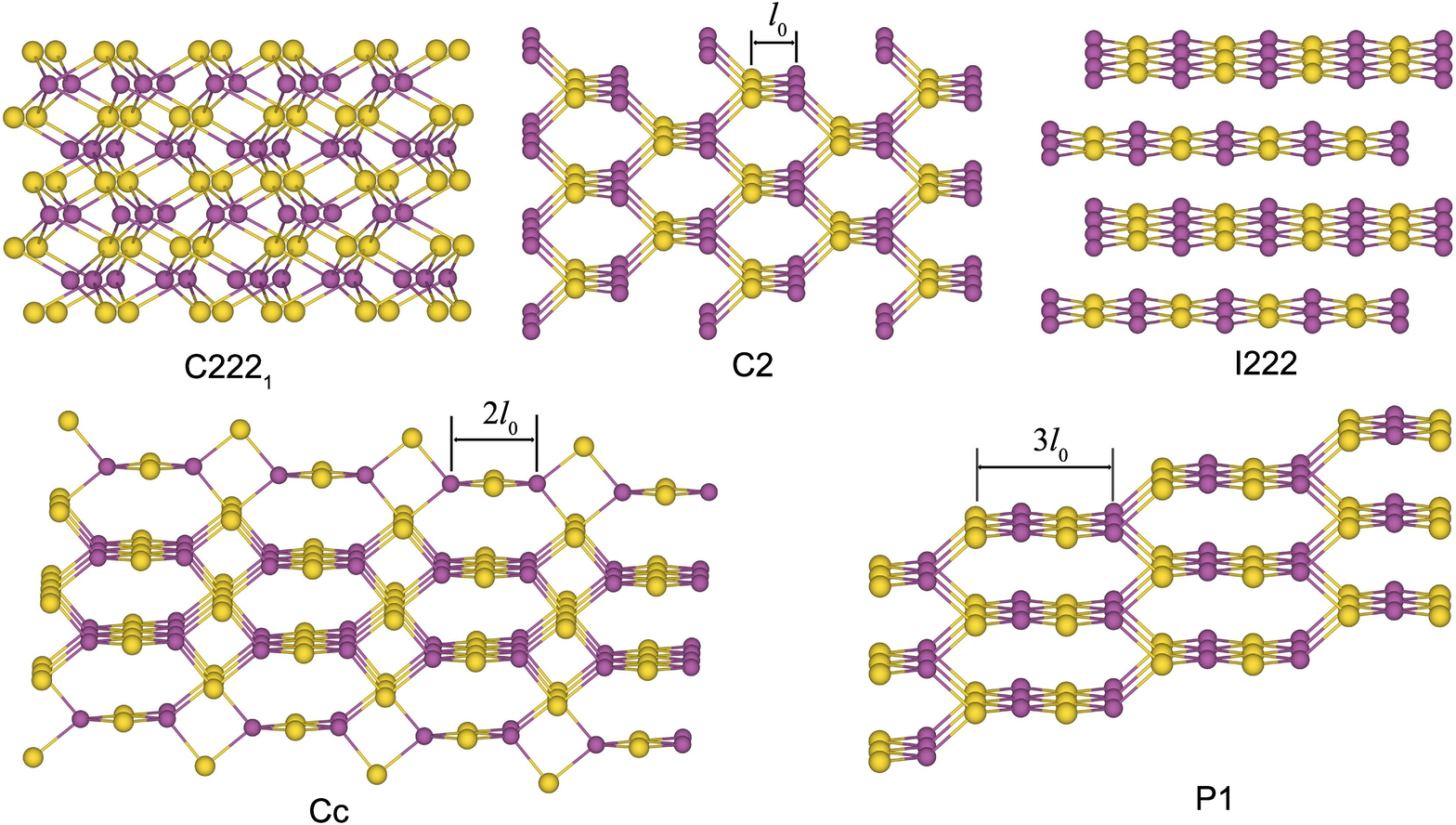}
  \caption{(Color online) Sodium-scandium framework of the examined low-energy structures of NaSc(BH$_4$)$_4$. Yellow and purple spheres represent sodium and scandium atoms. The new $C2$, $Cc$, $P1$, and $I222$ structures are characterized by hexagons which are clearly shown in the Figure. These structures are different in the length of the parallel sides, which is $l_0$, $2l_0$, $3l_0$, and infinite. Here $l_0\simeq 3.2$\AA~ is the sodium-scandium separation along the direction which is perpendicular to the line of view.} \label{fig:structure}
  \end{center}
\end{figure*}

The predicted structures are similar to the $C222_1$ structure in the geometry of the [Sc(BH$_4$)$_4$]$^-$ complex anions (see Fig. \ref{fig:coordination}). On the other hand, as shown in Figs. \ref{fig:coordination} and \ref{fig:structure}, the Na-Sc frameworks of the new structures and the $C222_1$ structure are completely different. In the novel structural motif, each Na (Sc) atom is coordinated to four Sc (Na) atoms at a distance of $\simeq 5$\AA. All four-fold coordinated atoms are arranged at the corners of either a rectangle with nearly equal sides or a tetrahedron, leading to  two types of Na-Sc ``bonds", either arranged on the planes of the rectangles mentioned above or interlinking them (see Fig. \ref{fig:coordination}). The Na-Sc planes are stacked in parallel planes and separated by a distance of $\simeq 6.1$\AA.  Viewed along these Na-Sc planes, the new structural motif is characterized by hexagons which have two parallel sides placed on the planes. The lengths of these parallel sides are $l_0$, $2l_0$, and $3l_0$ for the $C2$, $Cc$, and $P1$ structures (here, $l_0 \simeq 3.2$\AA~ is the Na-Sc separation in the direction perpendicular to the line of view --- see Fig. \ref{fig:structure}). For the $I222$ structure, this length is infinite since it is composed solely of the parallel Na-Sc planes.

A comparison of the XRD patterns of the new structures with both the $C222_1$ and $Cmcm$ phases clearly shows that the structures are distinct. All new structures are dynamically stable with no imaginary phonons in the whole Brillouin zone. The structural data, the phonon density of states and the simulated XRD patterns of the examined structures are given in the Supplemental Materials. \cite{supplement}

\begin{figure}[t]
  \begin{center}
    \includegraphics[width=8.25 cm]{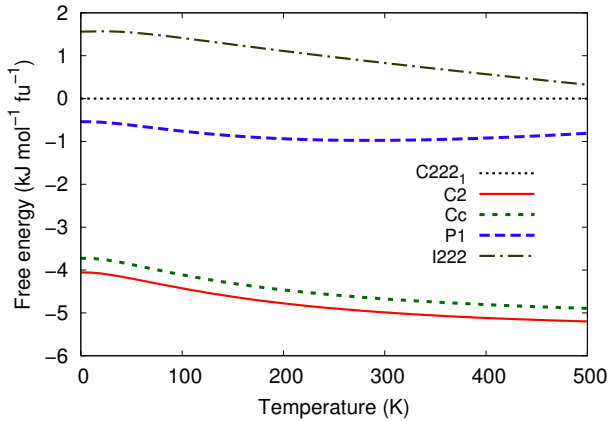}
  \caption{(Color online) Helmholtz free energies of the low-energy structures given with respect to the $C222_1$ structure.} \label{fig:free_energy}
  \end{center}
\end{figure}

We examine the stability of the predicted structures at finite temperatures by computing their Helmholtz free energies at the PBE level. In our calculations, vibrational free energies were determined within the harmonic approximation calculated from the phonon frequency spectrum. Fig. \ref{fig:free_energy} indicates that up to the melting temperature of ($\simeq$ 410K), the $C2$, $Cc$, and $P1$ are energetically favorable over the $C222_1$ structure. The $I222$ structure, on the other hand, is thermodynamically less stable than the $C222_1$ structure.

The formation of the predicted structures, which are energetically competing with the experimental $C222_1$ phase, may be possible. According to the Ostwald's step rule in crystal nucleation, \cite{Ostwald_rule} instead of forming the most stable phase directly from a solution, the system crystallizes in a step process where it transforms from less stable phases to phases of higher stability.  In particular, the phase transformation would first occur towards a phase that requires the least activation energy. Overall, nucleation and crystallization are complex processes that often depend on the synthesization method. Therefore, there are chances for the predicted structures exist, given that the suitable synthesization method conditions are chosen.

We then examine the electronic structures of the predicted phases of NaSc(BH$_4$)$_4$ by performing $\rm GW$ calculations. \cite{GW} The calculated energy band gaps $E_{\rm g}^{\rm GW}$ are shown in Table \ref{table_energy}. We found that the $Cmcm$ and $C222_1$ structures have band gaps of $8.10$ eV and $7.85$ eV, respectively. By referring to the result by Huang {\it et al.} \cite{Huang:NaSc}, the $\rm GW$ correction for these structures is roughly 3.0 eV. Similar values for the $\rm GW$ band gaps were also found for the predicted phases.

Several changes in the thermodynamic properties associated with the predicted low-energy structures of NaSc(BH$_4$)$_4$ can be qualitatively estimated. At the room temperature $T=298$ K, the reaction (\ref{Eq:decomp}) was predicted \cite{Kim12:NaSc, Kim13:NaSc} to occurs at the hydrogen pressure $P_{\rm H_2}$ from $10^3$ to $10^7$ bar. As the minimum hydrogen pressure $P_{\rm H_2}^{\rm min}=10^3$ bar was predicted for the $Cmcm$ phase, this parameter can be estimated for other low-energy structures of NaSc(BH$_4$)$_4$ assuming that there is no change in the products of the reaction (\ref{Eq:decomp}). Hence, the change of the Gibbs energy $\Delta G(T)$ is given by the change in the Helmholtz free energies, calculated at the PBE level and shown in Table \ref{table_energy}. From $P_{\rm H_2}^{\rm min}$, predicted for the $Cmcm$ phase by using the relation \ref{Eq:vHoff}, one arrives at $P_{\rm H_2}^{\rm min}=770$ bar for the $C222_1$ phase. Corresponding to the $C2$ phase, the $P_{\rm H_2}^{\rm min}$ can be reached at $546$ bar which is only half as much as the required pressure for the $Cmcm$ phase. Estimated $P_{\rm H_2}^{\rm min}$ of other structures are also given in Table \ref{table_energy}. As discussed in Sec. \ref{sec:intro}, it was also predicted \cite{Kim12:NaSc, Kim13:NaSc} that at $P_{\rm H_2}=100$ bar, the proposed reaction (\ref{Eq:decomp}) occurs at temperatures between 14 K and 223 K if NaSc(BH$_4$)$_4$ is in the $Cmcm$ phase. For the newly proposed phases, these temperatures are expected to change. Although a quantitative determination requires $\Delta G(T)$ to be fully computed according to Eq. (\ref{Eq:decomp}), one can qualitatively find from the van't Hoff plots reported \cite{Kim12:NaSc, Kim13:NaSc} that the temperature changes would be positive, i.e., the corresponding reaction temperatures are higher than those predicted for the $Cmcm$ phase, i.e., somewhat closer to the experimentally determined temperature ($>400$K). 

\section{Conclusions}
In summary, we show that the experimentally reported $Cmcm$ structure of NaSc(BH$_4$)$_4$ is dynamically unstable based on first-principles calculations. By exploring the unstable phonon modes, we predict a dynamically stable $C222_1$ structure which is lower in energy than the $Cmcm$ structure by 3.8 kJ mol$^{-1}$f.u.$^{-1}$ but explains the experimental XRD pattern equally well.

In addition to the $C222_1$ structure, we report a class of four low-energy structures predicted by the MHM for NaSc(BH$_4$)$_4$. These structures exhibit a new structural motif where each sodium (scandium) atom is coordinated to four scandium (sodium) atoms. Below the melting point reported for NaSc(BH$_4$)$_4$, all of these structures are energetically competing with the experimental $C222_1$ phase. According to the empirical Ostwald's step rule, the formation of the predicted structures may be possible. Once realized, the energy differences of these new structures may significantly influence the hydrogen pressure at which the dehydrogenation reaction (\ref{Eq:decomp}) of NaSc(BH$_4$)$_4$ occurs.

\begin{acknowledgments}
The authors thank Radovan \v{C}ern\'{y} for the experimental XRD data shown in Fig. \ref{fig:xrd} and many expert discussions. They thank Atsushi Togo, Tom\'{a}\v{s} Bu\v{c}ko, and Ki Chul Kim for valuable helps and comments. H. D. T., M. A., and S. G. gratefully acknowledge the financial support from the Swiss National Science Foundation. Computational work was performed at the Swiss National Supercomputing Center (CSCS) in Lugano. The space groups of the structures examined in this work were determined by {\sc findsym}, \cite{findsym} while Figs. \ref{fig:deform}, \ref{fig:coordination}, and \ref{fig:structure} were rendered with {\sc vesta}. \cite{vesta}
\end{acknowledgments}


\end{document}